\documentclass[letterpaper]{article}

\usepackage{parskip}
\usepackage{graphicx}
\usepackage{amsmath}
\usepackage{caption}
\usepackage{chngpage}
\usepackage{amssymb}
\usepackage{textgreek}
\usepackage [english]{babel}
\usepackage [autostyle, english = american]{csquotes}
\usepackage{multirow}
\MakeOuterQuote{"}

\begin{document}
\title{"Banking Deserts," City Size, and Socioeconomic Characteristics in Medium and Large U.S. Cities}
\author{Scott W. Hegerty\\
Department of Economics\\
Northeastern Illinois University\\
Chicago, IL 60625\\S-Hegerty@neiu.edu
}
\date{\today}
\maketitle

\begin{abstract}
A lack of financial access, which is often an issue in many central-city U.S. neighborhoods, can be linked to higher interest rates as well as negative health and psychological outcomes. A number of analyses of "banking deserts" have also found these areas to be poorer and less White than other parts of the city. While previous research has examined specific cities, or has classified areas by population densities, no study to date has examined a large set of individual cities. This study looks at 319 U.S. cities with populations greater than 100,000 and isolates areas with fewer than 0.318 banks per square mile based on distances from block-group centroids. The relative shares of these "deserts" appears to be independent of city population across the sample, and there is little relationship between these shares and socioeconomic variables such as the poverty rate or the percentage of Black residents. One plausible explanation is that only a subset of many cities' poorest, least White block groups can be classified as banking deserts; nearby block groups with similar socioeconomic characteristics are therefore non-deserts. Outside of the Northeast, non-desert areas tend to be poorer than deserts, suggesting that income- and bank-poor neighborhoods might not be as prevalent as is commonly assumed.
\end{abstract}

\section{Introduction}
Financial access is crucial for individuals who wish to improve their lives and businesses that aim to grow and expand. Yet as a growing share of this access has gone digital, many people have been left behind. A number of physical bank branches have closed as mobile banking has expanded, posing challenges for many older Americans and residents of central cities. Beyond its immediate economic impact, including higher interest rates due to poorer information on local neighborhood conditions, there are other negative effects as well. As Hegerty (2020) notes, a paucity of nearby bank branches not only signals a lack of investment by key community businesses, it also might thwart residents' abilities to fully integrate with the official financial system. According to the "spatial void" hypothesis described by Smith et al. (2008), "fringe" banks such as payday loan providers might step in to provide financial services at a higher cost to consumers. As a result, living in a banking desert might hurt community financial, and even physical and psychological, health (Eisenberg-Guyot et al., 2018). 

The economic impact of living in a banking desert can be significant. In a study of Ohio, Ergungor (2010) finds that interest rates are higher in areas with limited bank presence. Nguyen (2019) concludes that a reduction in the number of bank branches results in fewer loan originations. Information asymmetries may play a crucial role; particularly if outside lenders have less understanding of a community in which they lend or if distance increases monitoring costs for lenders (Degryse and Ongena, 2005). 

A number of studies have investigated bank access for a variety of spatial units and target areas. Brennan et al. (2011) find evidence that alternative financial providers serve poor neighborhoods where banks and credit unions are not located. On the contrary, Chen et al. (2014) do not find such support in a sample of U.S. counties. Important socioeconomic covariates with bank access include race, income, education, and housing tenure (Burkey and Simkins, 2004; Wheatley, 2010; Cover et al., 2011; Hegerty, 2016; and Dunham et al., 2018), although significant contributors vary by place.

Many of these studies are at the ZIP code or county level of analysis, which captures overall trends. Important city level studies include Hegerty (2016) for Buffalo and Milwaukee; Dunham et al. (2018) for Los Angeles, Las Vegas, and Miami; and Hegerty (2020) for Chicago. The impact of socioeconomic factors, particularly race, is shown to differ from city to city. There is increasing evidence that low-income banking deserts might be a strictly urban problem, limited particularly to the case of areas of large cities.  

At the national level, Kashian et al. (2018) study U.S. census tracts and, after controlling for populaton density, examine urban, rural, and suburban areas individually. Only in this first group is poverty shown to be negatively associated with the proximity to the nearest bank. Hegerty (2022) examines all census tracts in the contiguous United States along a range of population densities, and finds that the link between economic deprivation and limited bank access is strongest in tracts with relatively high densities that are comparable to Chicago's. Both studies, using different methods, find that an urban "banking desert" typically has fewer than 0.4 banks per square mile.

So far, no study has conducted a comprehensive analysis of urban bank access for the entire United States. This study does so, by calculating bank densities for all block groups in 319 U.S. cities with populations greater than 100,000. After isolating "banking deserts" based on a density criterion, the link between city size and the share of cities' bank deserts and differences within cities are examined. Overall, there is surprisingly little evidence of significant differences between the size of banking deserts along the range of city size, or of poverty or racial characterstics between deserts and "non-deserts." One exception might be in the U.S. Northeast.

\section{Methodology}

Coordinate data for bank locations in the entire lower 48 states and the District of Columbia, from the FDIC’s Summary of Deposits database as of June 30, 2019, are used for this analysis. These are mapped using the R package \textit{sf} and compared against U.S. census block groups. Using U.S. census data, 319 cities are identified as having more than 100,000 people as of 2019. These make up the study sample, although particular attention is paid to the largest cities, with populations above 250,000. 

Within each city (using census "Places" shapefiles), block groups are selected and buffers with 1- and 2-mile radii around each centroid are created. The number of banks are then counted within each buffer. Using the (slightly modified) criterion put forth by Hegerty (2020), a "banking desert" is defined if a block group has one or fewer banks within one mile or four or fewer banks within two miles. Since a 2-mile circle has four times the area of a 1-mile circle, both measures result in a density of (1/\textpi)   or 0.318 banks per square mile. Allowing for the larger radius eliminates the possibility where a single branch in a high-need neighborhood improves the perception of a neighborhood's financial access without producing measurable changes.  

These bank counts are then used to calculate a number of city-level measures. First, because larger cities might be expected to have more banks overall, median counts per city are compared across city size. Additional quantiles are also computed, particularly since the 5\% or 10\% thresholds might be used to classify a "banking desert" (as in Kashian, 2018). \ It might be expected that a city with a large number of poor people might have relatively few banks in its highest-poverty, as well as its lowest-poverty, block groups. 

Next, the relative size of banking deserts within each city (by population, area, and number of block groups as shares of the total) are compared. This can be used to assess wthat a "typical" banking desert might look like. It can also be used to isolate cities with exceptionally limited financial access. After comparing some numerical results with a visual analysis of certain representative cities, correlations are calculated between desert size and three key socioeconomic variables: the poverty rate, the percentage Black, and the percentage White. This is done across the range of city sizes, for both percentages and concentrations.

 For each city, the poverty \textit{rate} in its banking deserts is first calculated as 100 x (Desert population in poverty/Total desert population). The poverty \textit{concentration} is calculated as 100 x (Desert population in poverty/City population in poverty). The same is done for the other two variables and for all variables in non-desert areas as well. Here, calculating rates is useful because it might be expected that deserts are poorer and have more Black and fewer White residents than non-deserts, while concentrations can help show whether banking deserts have disproprtionately more (or fewer) of a city's poor or non-White residents. Correlations and bivariate regressions between desert size and each of these six variables are then plotted, with a distinction between "medium" (100-250K) and "large" (greater than 250K) residents. This distinction might help show whether banking deserts are indeed mainly a "big city" phenomenon. 

Are banking deserts in large cities poorer or less White than in smaller cities? To answer this question, rolling quantiles are calculated for the six socioeconomic variables described above. Beginning with the 10 largest cities, the 10\%, 50\%, and 90\% thresholds are depicted for both desert and nondesert areas. The sample is then expanded, one by one, until the values are calculated for all 319 cities. 

For example, median poverty (the 50\% quantile) in the two types of area can be directly compared for any subset of the largest cities, up to the entire sample. It may be expected that banking deserts have higher average poverty rates than other parts of the city. The same might be said for extreme poverty, while it is possible that non-deserts might have better poverty rates at all quantiles. At the same time, there might be a different relationship in a city of 100,000 than there is in a city with 500,000 residents.  Similarly, the median percentages of Black or White residents might be expected to be higher or lower, respectively, than elsewhere in the city mig; thisht not hold for all city sizes. The same can be said for population concentrations, i.e. where banking "deserts" might have large shares of a city's poor residents, but not necessarily in every city.  

Finally, the distributions of these six variables in deserts and non-deserts are directly compared for all six variables. It might be expected that banking deserts might not only have higher median poverty, but also more instances of high poverty rates. Likewise, the average difference in the sample's poverty differential (measured as desert minus nondesert poverty rates) is expected to be positive. This relationship might also vary by city size, with greater differentials in larger cities. These relationships, like many of those that are hypothesized above, do not seem to be the case. In fact, there appears to be little statistical difference, either between deserts and nondeserts, or between smaller and larger cities outsideof the U.S. Northeast. This might have implications for the larger discussion of banking access in the United States. The results are described below.

\section{Results}
The first set of analyses focuses on the relationship between bank access and city size. Figure 1 presents median bank counts for each city, calculated using 1- and 2-mile buffers from all block group centroids. As expected, there are more banks within a 2-mile circle, but the values are closely connected (with a correlation coefficient of 0.902). There is no evidence of any relationship between city size and median bank counts, however, with a range of values among the large number of "small" cities. Excluding New York (with a large number of banks and the largest population), there is also no evidence of cities with populations greater than one million people having higher median bank counts than cities below this threshold.

\begin{figure}[!htbp]
\hfill

\caption{
Distribution of Median Bank Counts, By City Size.}
\includegraphics[width=1\textwidth]{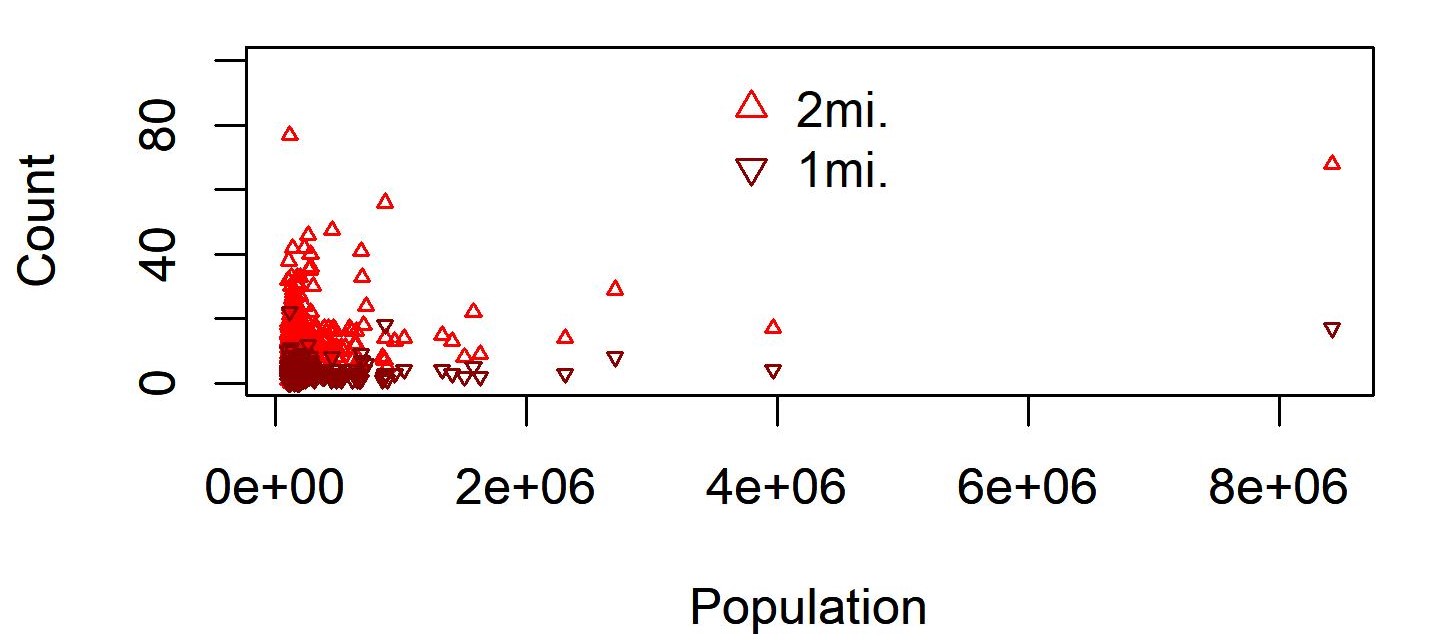}
\caption*{\footnotesize
Median counts calculated for given radii around all block group centroids within each city.\\
Correlations between population and median count at 1 and 2 miles = 0.266 and 0.291 (Pearson) and -0.095 and 0.014 (Spearman), respectively.\\
Correlation between 1-mile and 2-mile medians = 0.902.\\
}
\end{figure}

The distributions of median values for all 319 cities are plotted in Figure 2. Half of the cities have, on average, at least 3 banks within one mile or 18 banks within two miles, while the cities with the worst financial access have no banks within a mile or fewer than four banks within two miles in half their tracts. Because these cities might have the highest poverty rates--or even the largest numbers of poor residents--correlations between these variables and the quantile values are plotted in the lower panel. There is, however, almost no bivariate relationship. Cities with low median bank counts are not likely to be poorer than others. 

\begin{figure}[h]
\begin{center}
\hfill

\caption{
Densities of Median Bank Counts and Correlations With Poverty.}
\includegraphics[width=.8\textwidth]{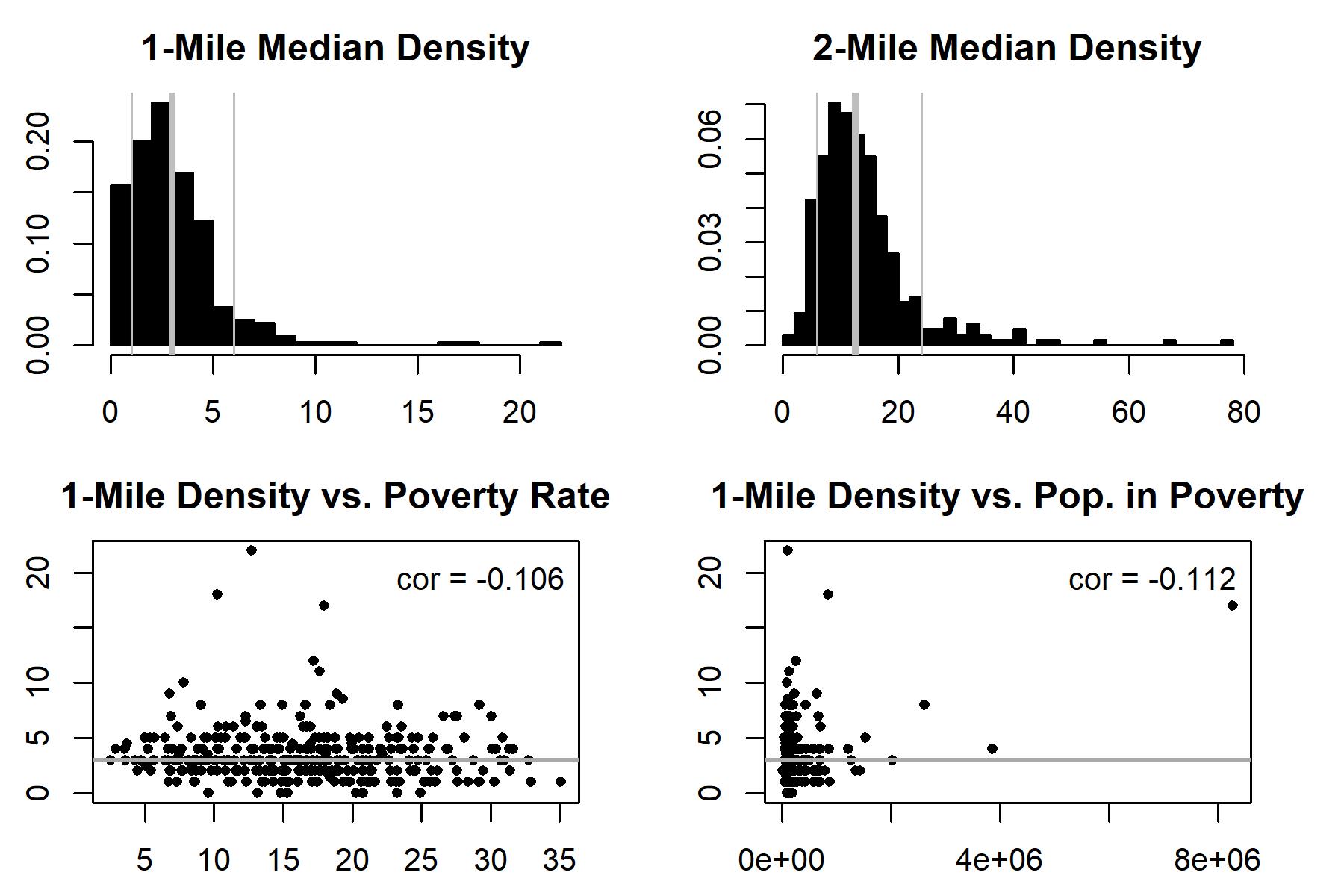}
\end{center}
\caption*{
\footnotesize
Median counts calculated for given radii from all block group centroids within each city.\\
Top panel: Vertical lines = 10\%, 50\%, and 90\% quantile values.\\
Bottom panel: Bivariate regression line in gray. N = 319 cities,
}
\end{figure}

\begin{figure}[h]
\begin{center}
\hfill

\caption{
Proportion of "Banking Deserts" by City.}
\includegraphics[width=.8\textwidth]{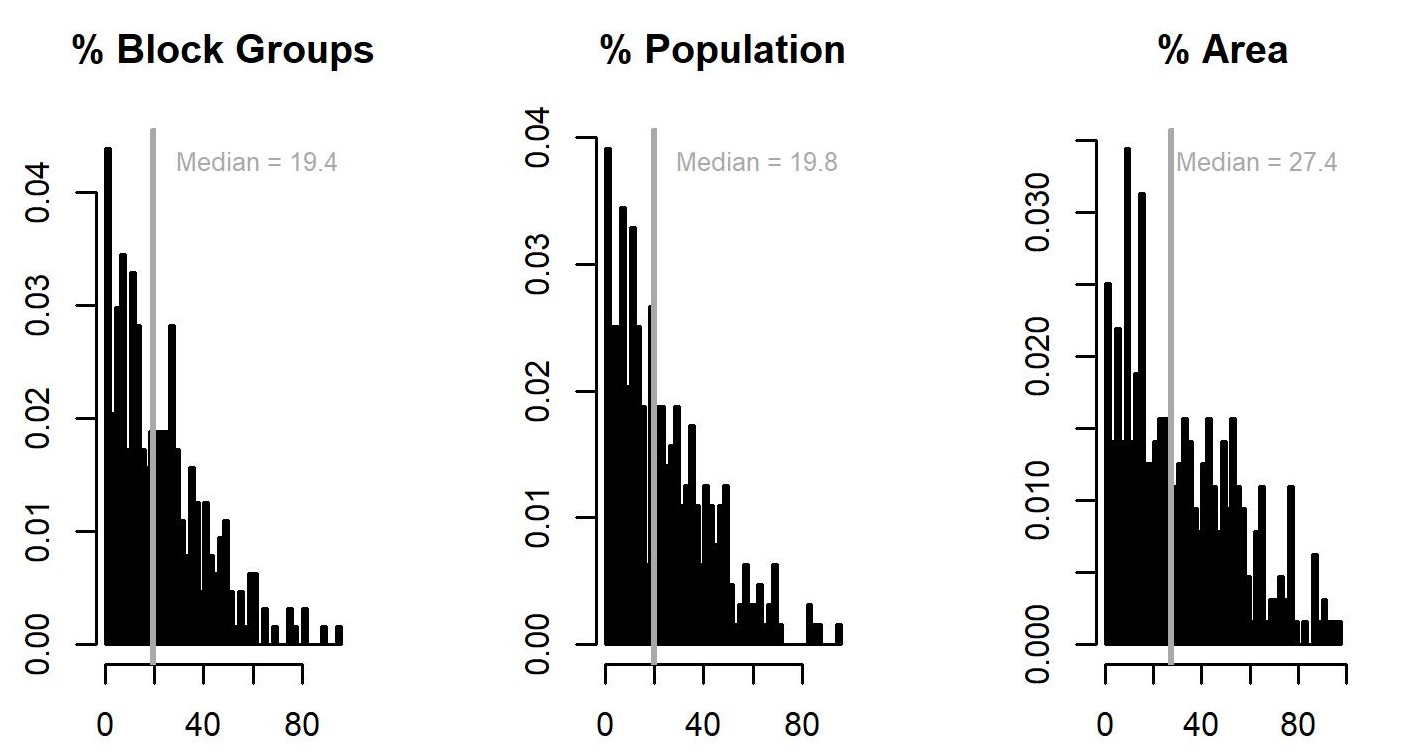}
\end{center}
\caption*{
\footnotesize
Percentages located within defined "banking deserts" in N = 319 cities.\\
Gray vertical line = Full-sample median values.
}
\end{figure}

Next, the relative size of banking deserts in the sample of 319 cities is examined. Figure 4 shows that the median size of urban banking deserts is about 20\%, in terms of population and number of block groups, land area, and about 27\% when measured in terms of land area. The modal share of banking deserts is low, however; the distributions are heavily skewed. 

What do banking deserts look like spatially? Figure 4 shows the example of Milwaukee; block groups that are identified as banking deserts are outlined. The choropleth maps show that these North Side areas are poorer and have higher proportions of Black residents than other parts of the city, while the Northwest Side is more diverse. But many poor and/or majority Black neighborhoods cannot be classified as "banking deserts." As a result, it is quite possible that non-deserts might be as poor or poorer than deserts, leading to little distinction between areas. This can be the case for any city.

\begin{figure}[h]
\hfill

\caption{"Banking Deserts" in Milwaukee and Three Comparison Cities.}
\begin{center}
\includegraphics[width=.22\textwidth]{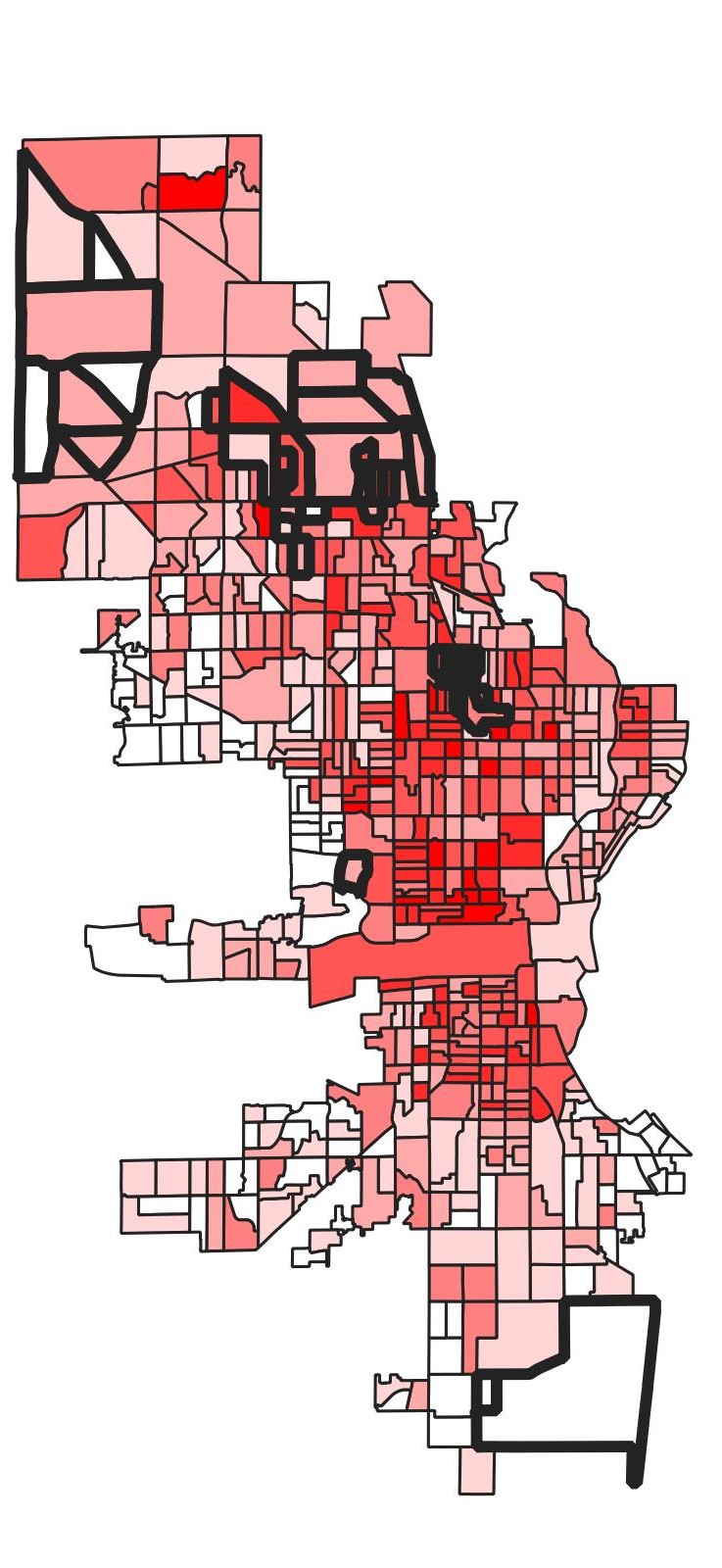}
\includegraphics[width=.21\textwidth]{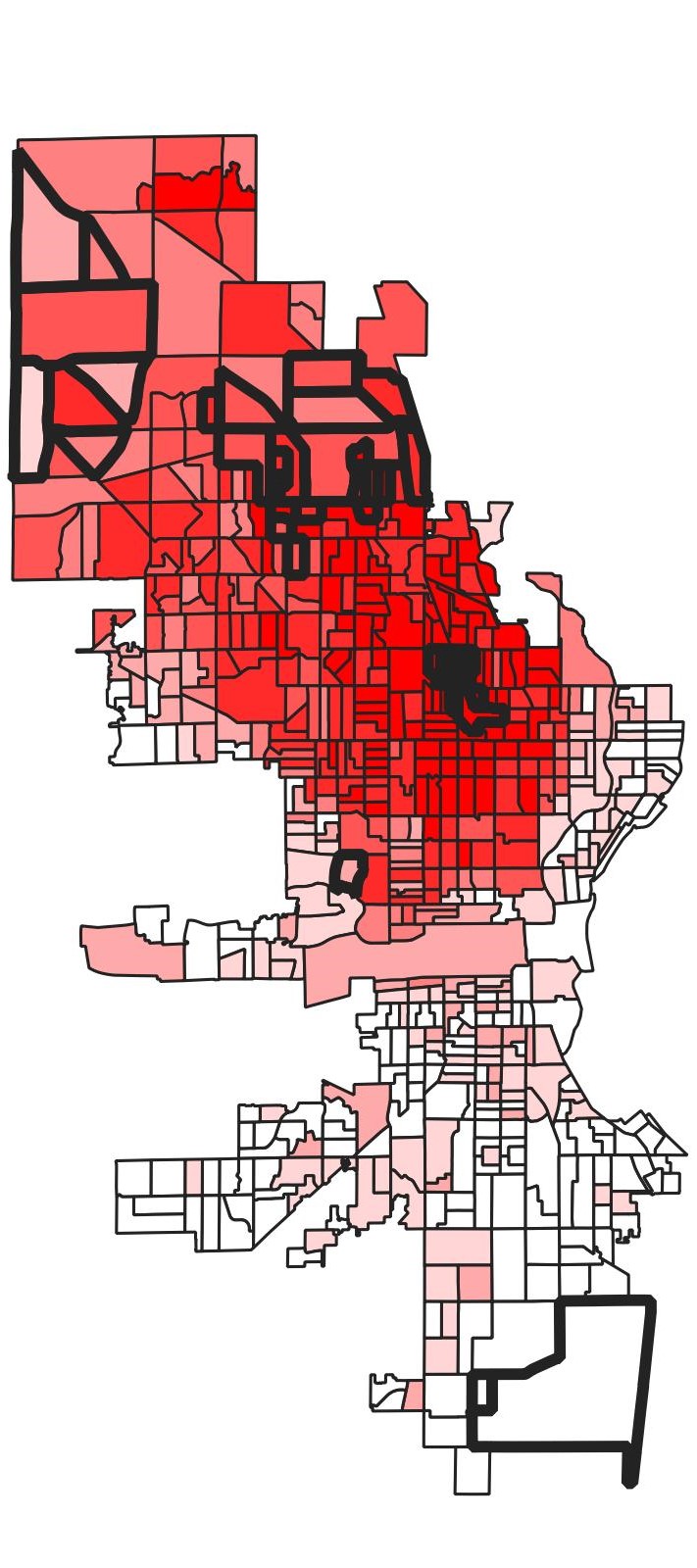}
\includegraphics[width=.22\textwidth]{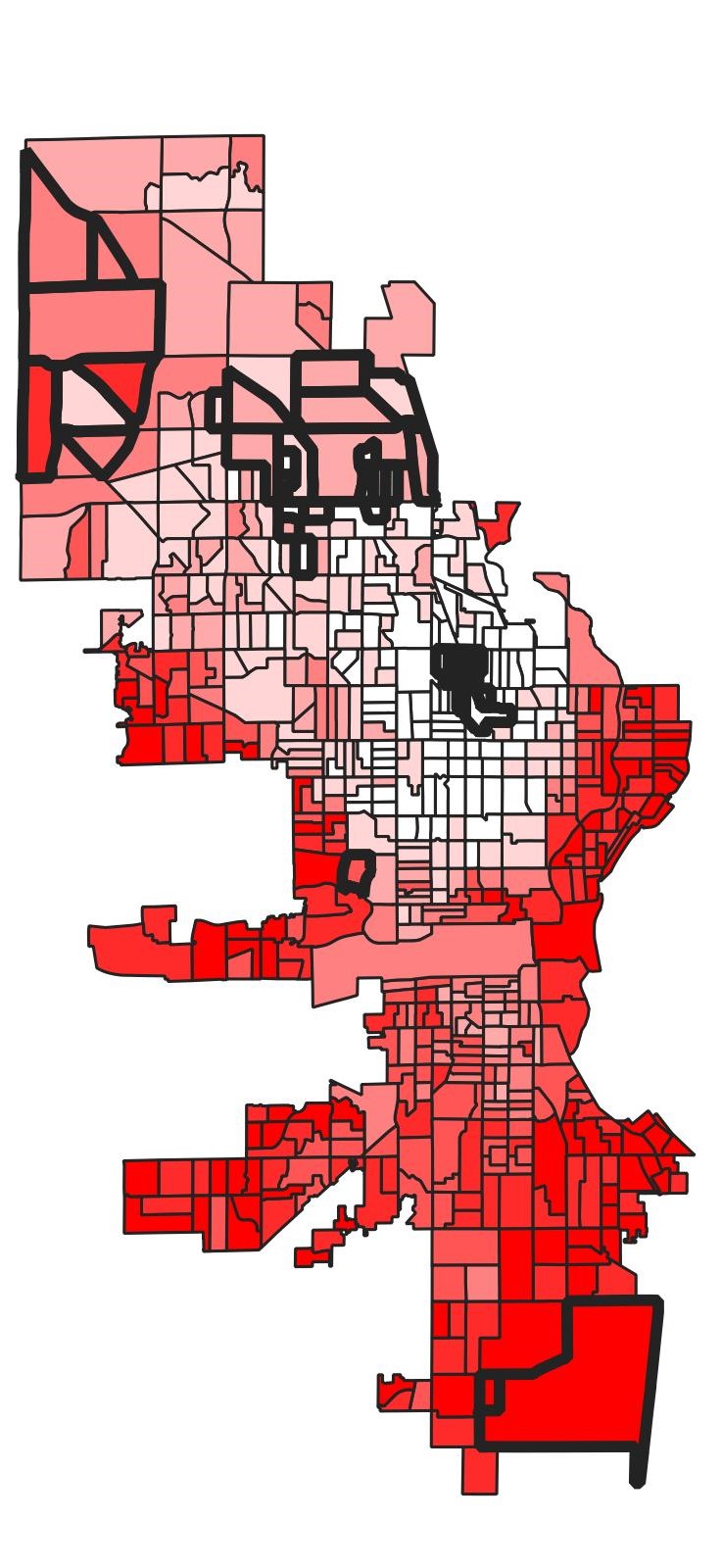}

\includegraphics[width=.21\textwidth]{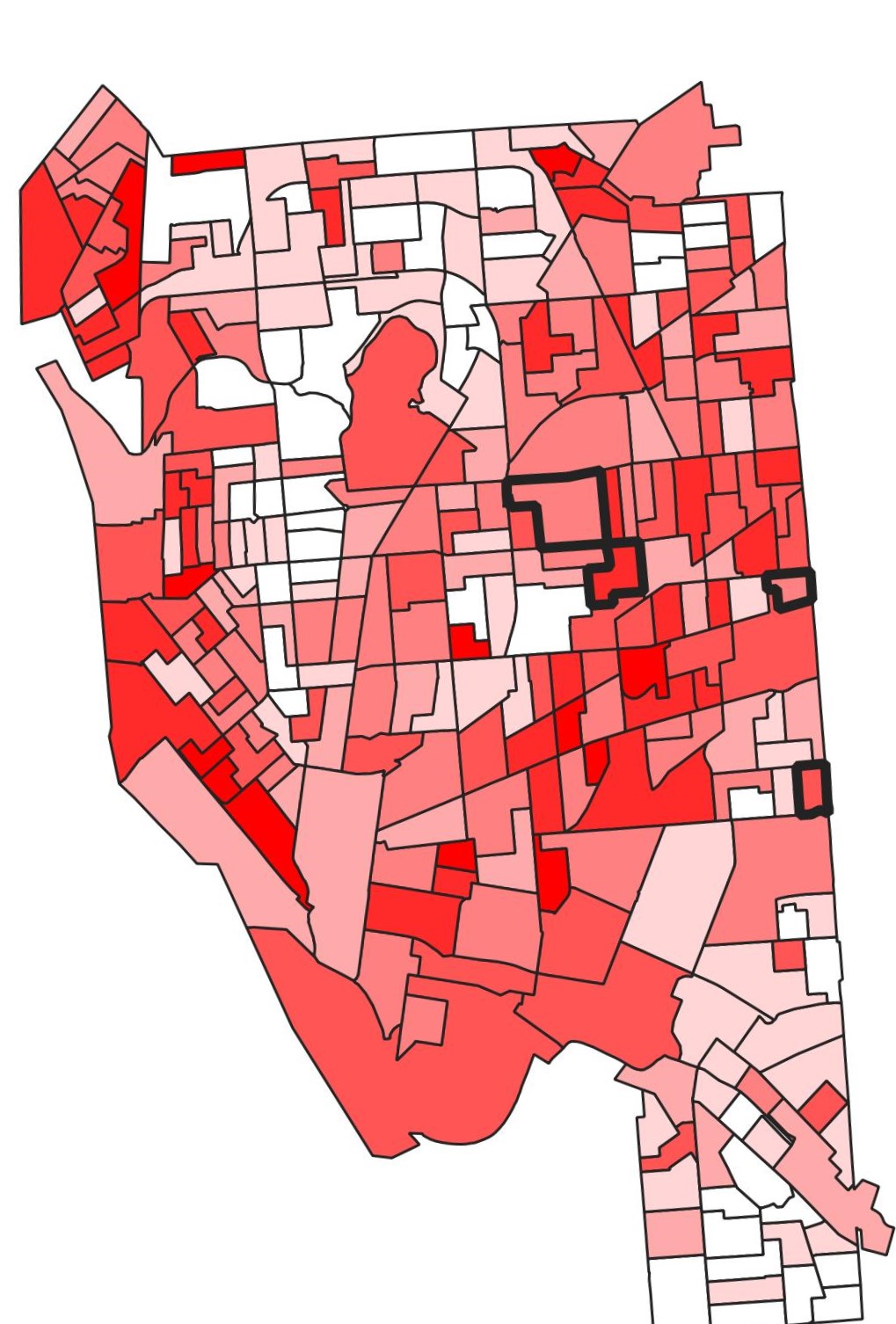}
\includegraphics[width=.2\textwidth]{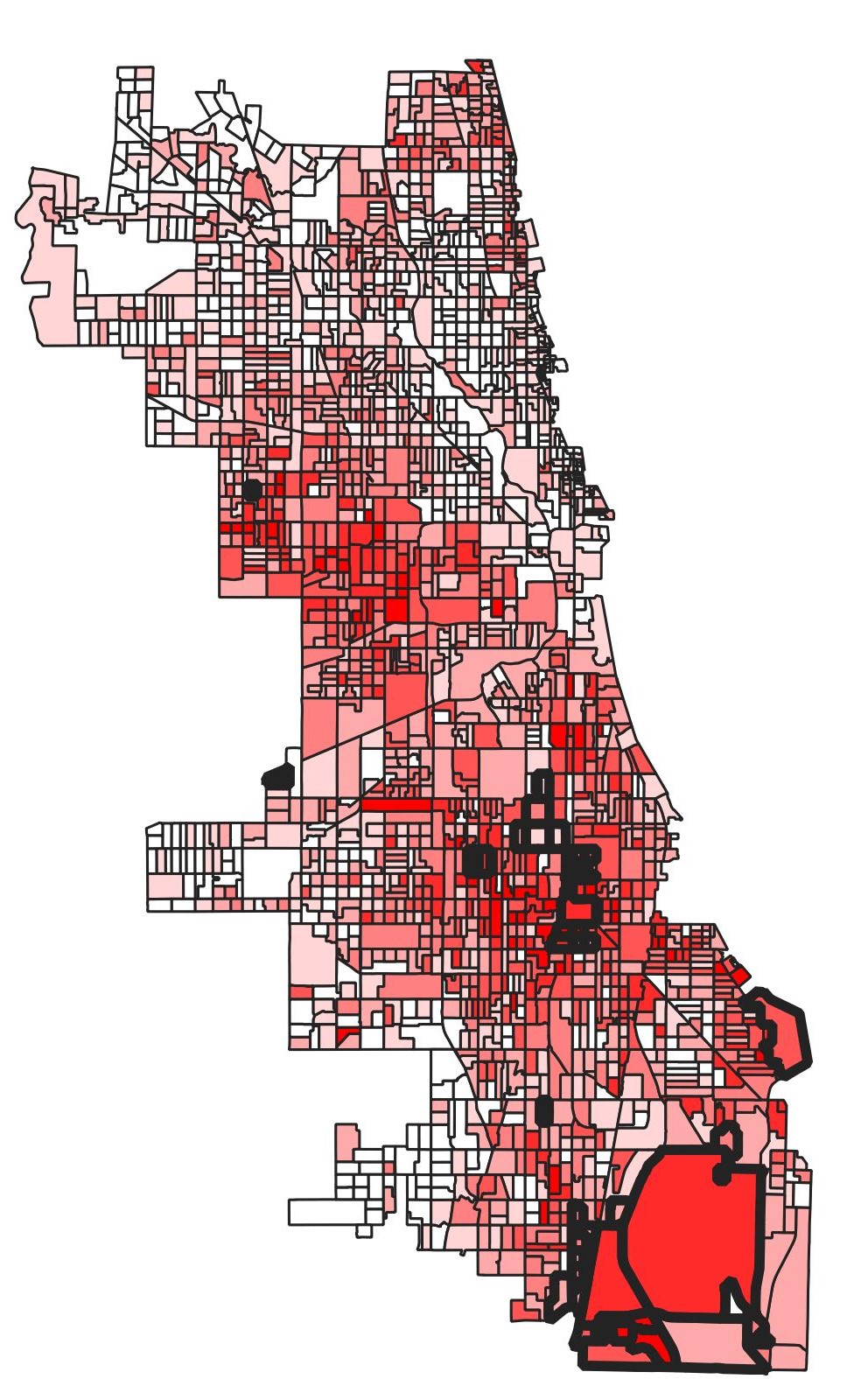}
\includegraphics[width=.25\textwidth]{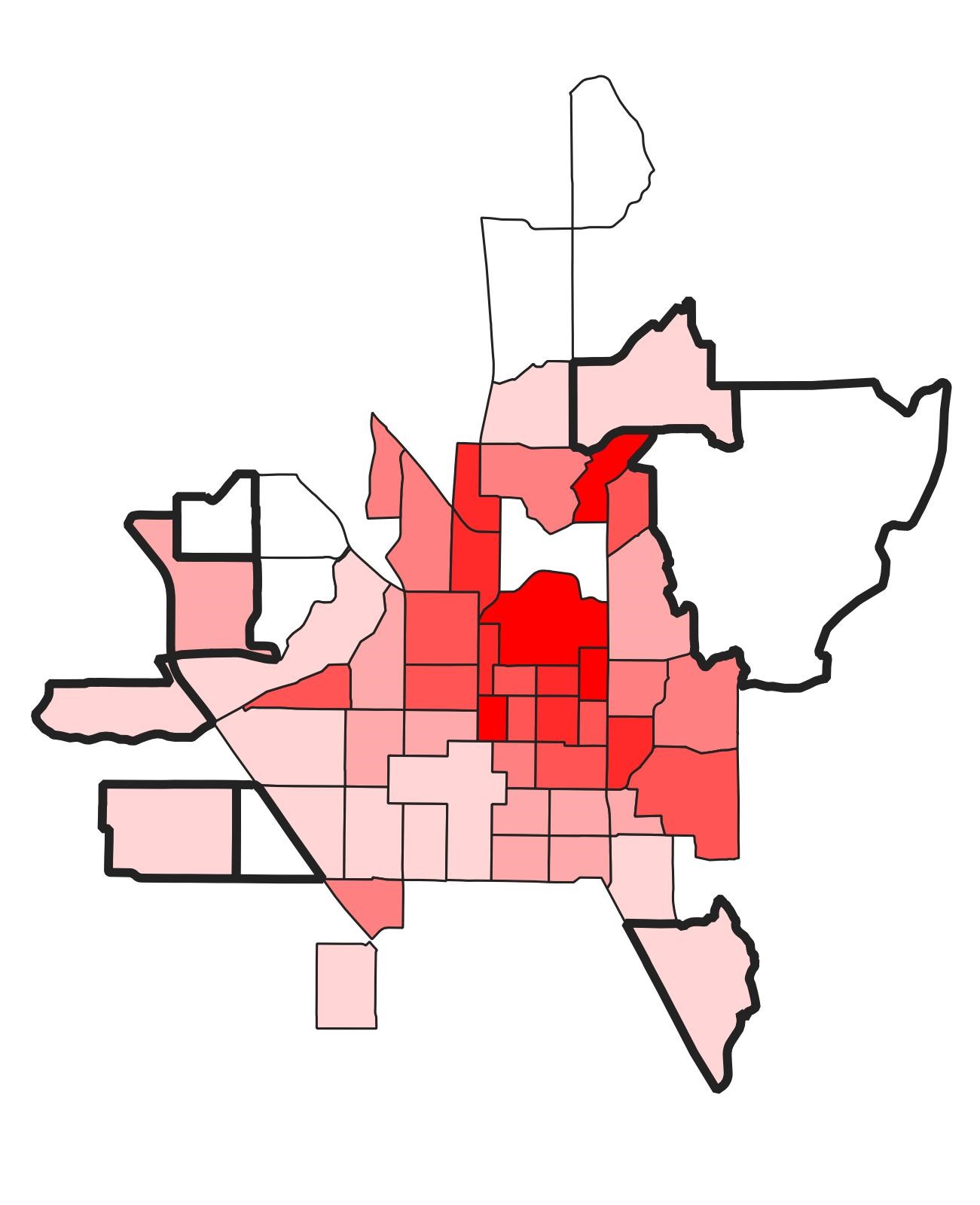}
\caption*{
\footnotesize
Top panel: Milwaukee poverty rate, percentage Black, percentage White.\\
Bottom panel: Poverty rates in Buffalo, Chicago, Provo (Utah).\\
Darker = higher rates for each variable.\\
Black outlines: "Banking desert" block groups.\\ 
}
\end{center}
\end{figure}

The lower panel shows the presence of hgh-poverty block groups among the banking deserts on Buffalo's East Side and Chicago's West and South Sides. Provo (Utah), on the other hand, is an entirely different type of city--with its banking deserts located on its outskirts. This is likely due to regional variations in city types; in the South and West, cities are often less dense. Many are also "overbounded," containing areas that would be considered to be suburban in the Northeast. This confirms the plots above that show that poor financial access is not necessarily linked to high poverty rates.

Poverty rates, as well as the other five key variables, are plotted against banking desert size in Figure 5. Here, the cities' percentage White is negatively correlated with the share of population living in banking deserts. The other variables show positive relationships, although the rates' correlations are much smaller than those of the concentrations. Cities with larger deserts larger concentrations of all types of people living in these deserts. Both the Black and White concentrations have large positive correlations with desert size. 

\begin{figure}[h]
\begin{center}
\hfill

\caption{Bivariate Relationships With Desert Proportions, by City Size.}
\includegraphics[width=.8\textwidth]{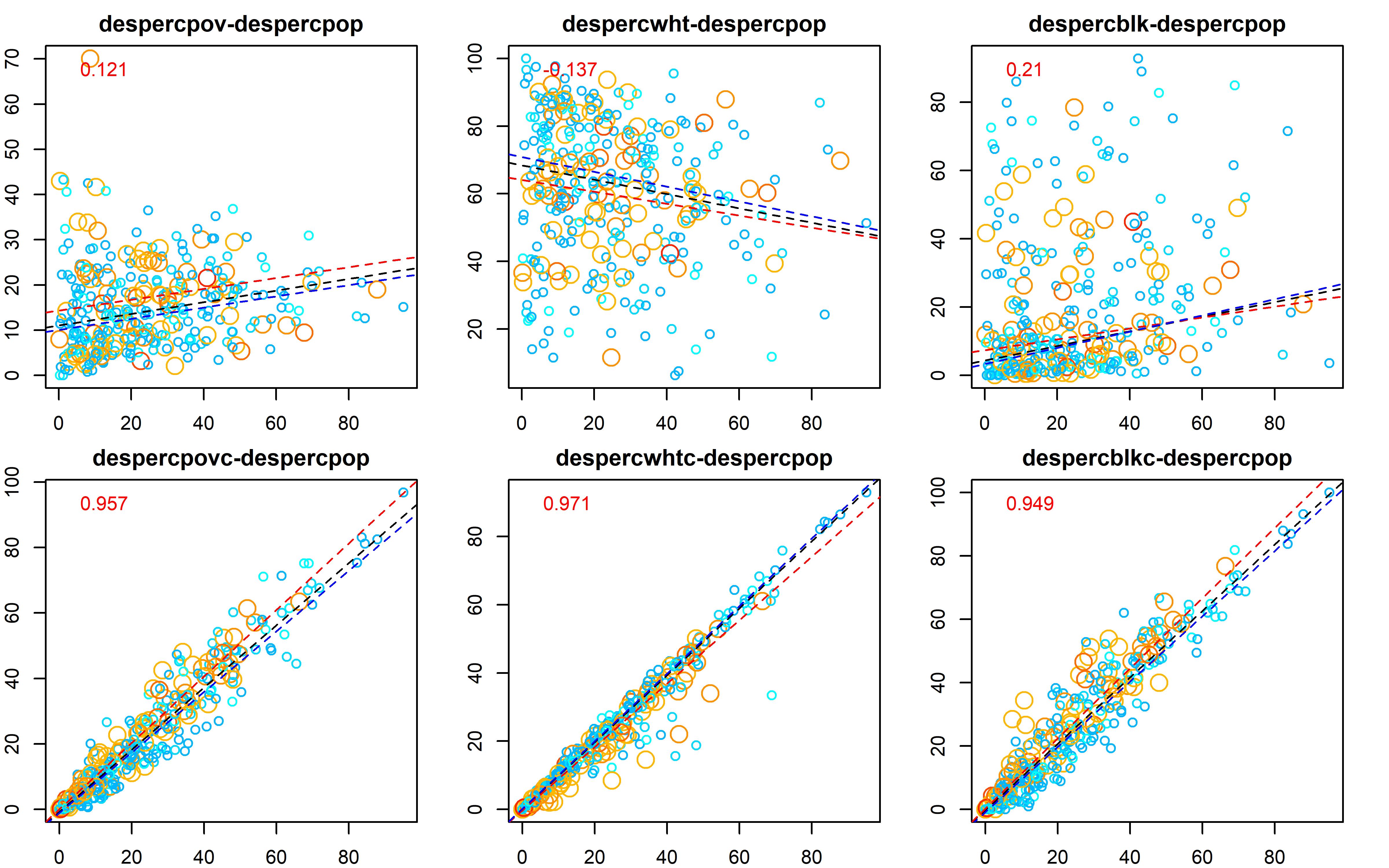}

\caption*{
\footnotesize
Spearman correlations in upper left corner.\\
Larger, red circles = cities with populations above 250000.\\
Smaller, blue circles = cities with populations below 250000.\\
All plots labeled as y-x axes.\\
Dashed lines = nonparametric (Thiel-Sen) regression lines.\\	
Full sample = black, large cities = red, smaller cities = blue.
}
\end{center}
\end{figure}

\begin{table}[h]
    \begin{adjustwidth}{-.5in}{-.5in}  
\caption{Thiel-Sen Bivariate Regression Results.}
        \begin{center}
\footnotesize
\textit{DV = \% Population in "Banking Desert" block groups within each city}
\begin{tabular}{lrrr}
&Full&Large&Small\\
\hline
Constant&11.017 (0.000)&14.356 (0.000)&10.121 (0.000)\\
despercpov&0.128 (0.000)&0.113 (0.023)&0.123 (0.000)\\
\hline
Constant&63.385 (0.000)&59.918 (0.000)&63.916 (0.000)\\
despercwht&-0.061 (0.157)&-0.079 (0.085)&-0.050 (0.610)\\
\hline
Constant&6.992 (0.000)&7.357 (0.000)&7.034 (0.000)\\
despercblk&0.061 (0.003)&-0.006 (0.635)&0.077 (0.000)\\
\hline
Constant&5.729 (0.000)&8.517 (0.000)&4.638 (0.000)\\
despercpovc&0.52 (0.000)&0.429 (0.000)&0.549 (0.000)\\
\hline
Constant&6.434 (0.000)&9.174 (0.000)&5.765 (0.000)\\
despercwhtc&0.471 (0.000)&0.28 (0.000)&0.516 (0.000)\\
\hline
Constant&5.536 (0.000)&12.953 (0.000)&3.849 (0.000)\\
despercblkc&0.586 (0.000)&0.404 (0.000)&0.621 (0.000)\\
\hline
\end{tabular}
\end{center}
\end{adjustwidth}

\caption*{
\footnotesize
Independent variables are listed in each row and are calculated at the city level.\\
"Large" cities: Population greater than 250,000 (N = 83);\\ "Small" cities = all others (N = 236).
P-values in parentheses.
}
\normalsize

\end{table}

\begin{figure}[h]
\hfill
\begin{center}
\caption{Rolling Quantiles (10\%, 50\%, 90\%) by City Size.}
\includegraphics[width=.4\textwidth]{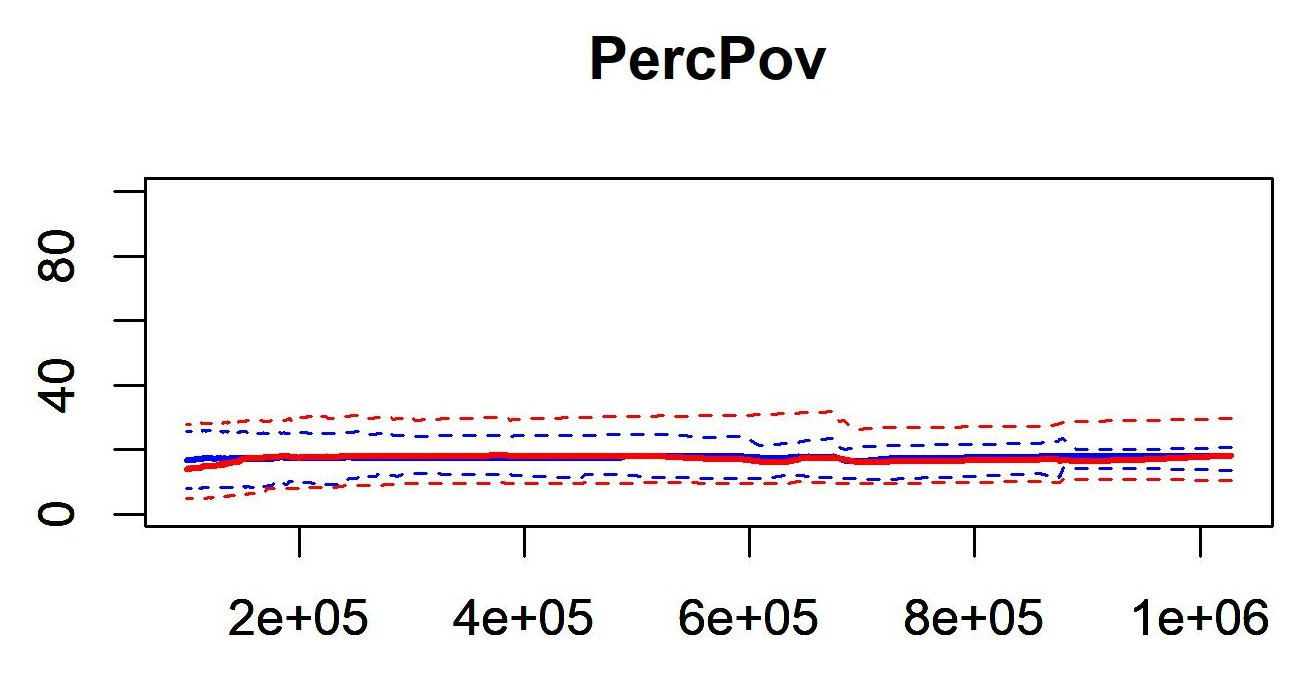}
\includegraphics[width=.4\textwidth]{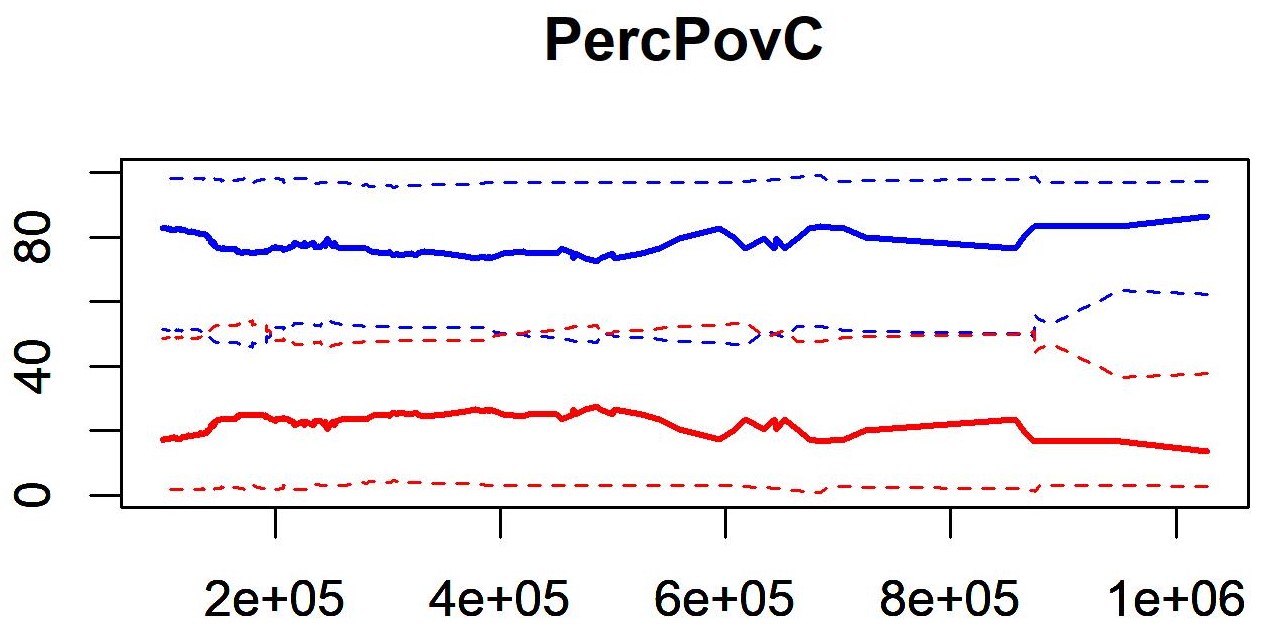}

\includegraphics[width=.4\textwidth]{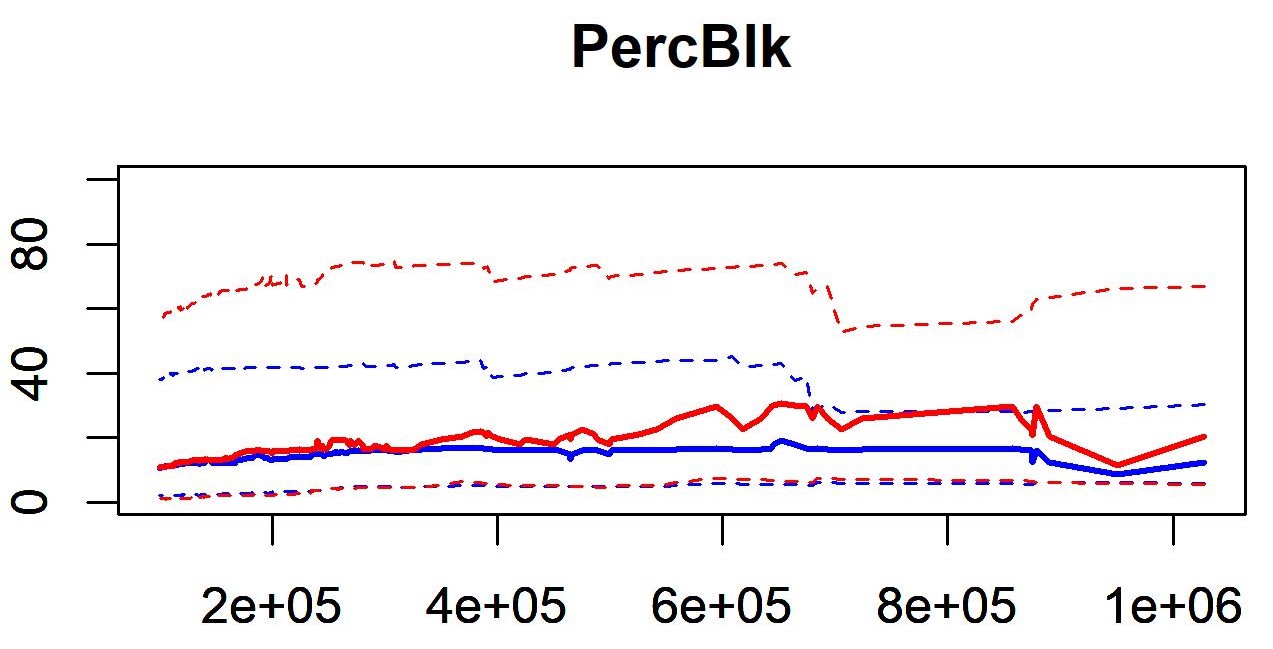}
\includegraphics[width=.4\textwidth]{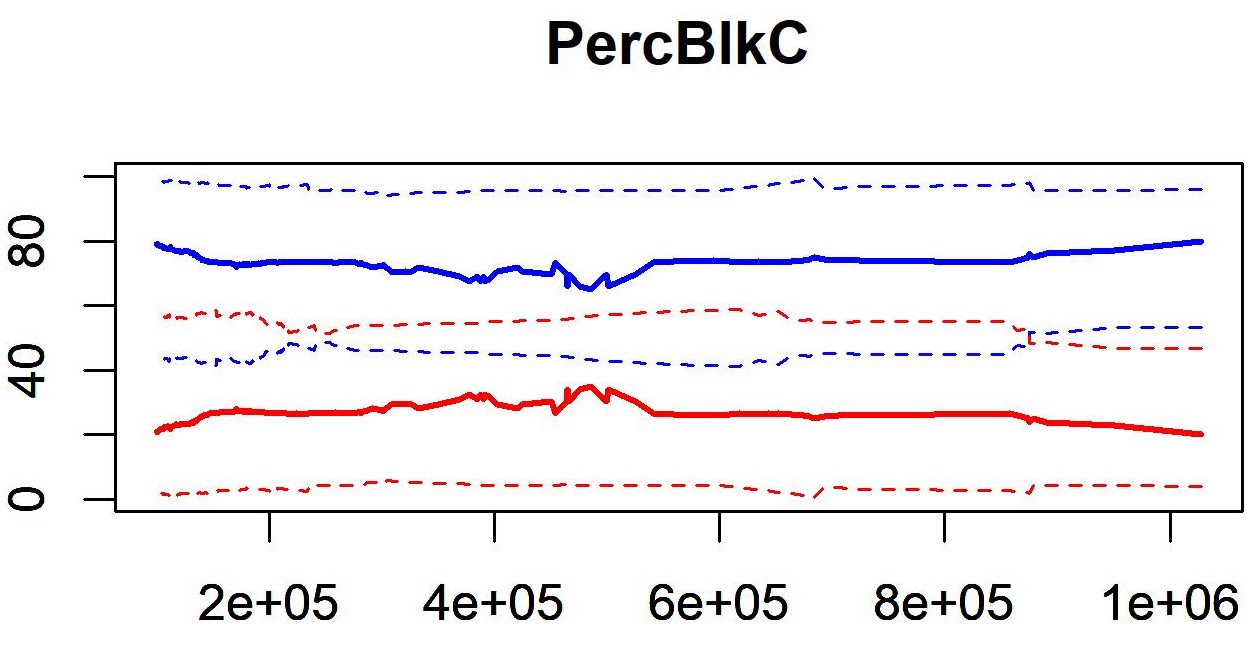}

\includegraphics[width=.4\textwidth]{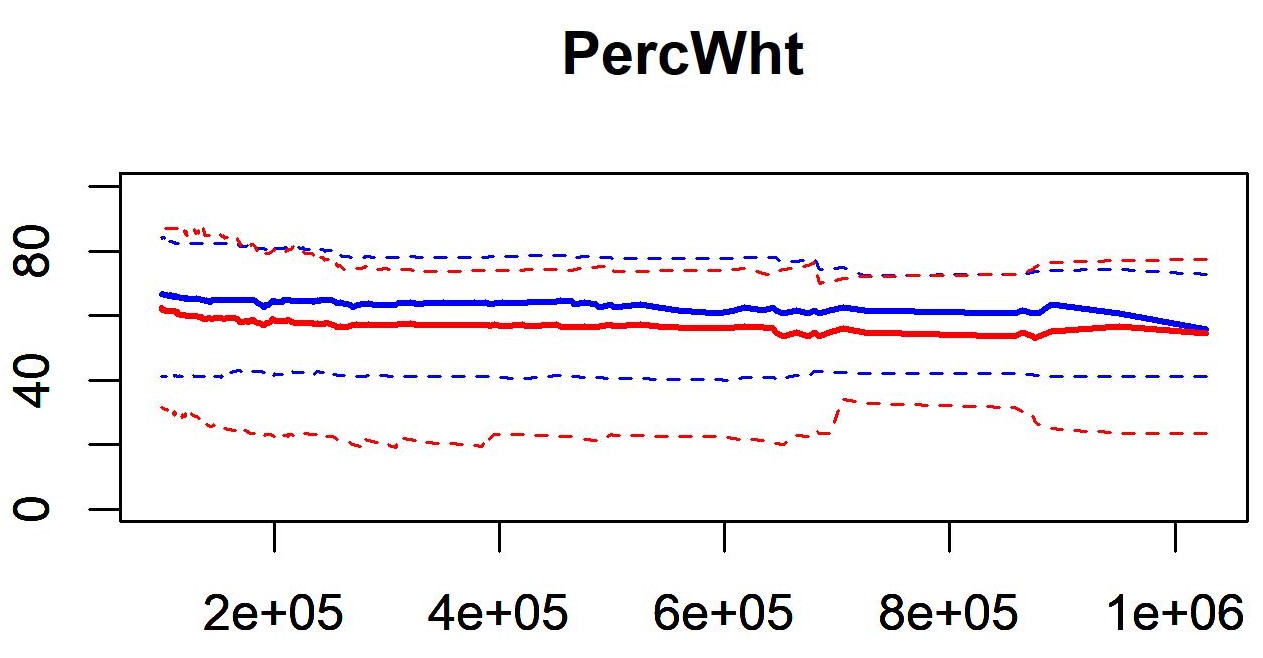}
\includegraphics[width=.4\textwidth]{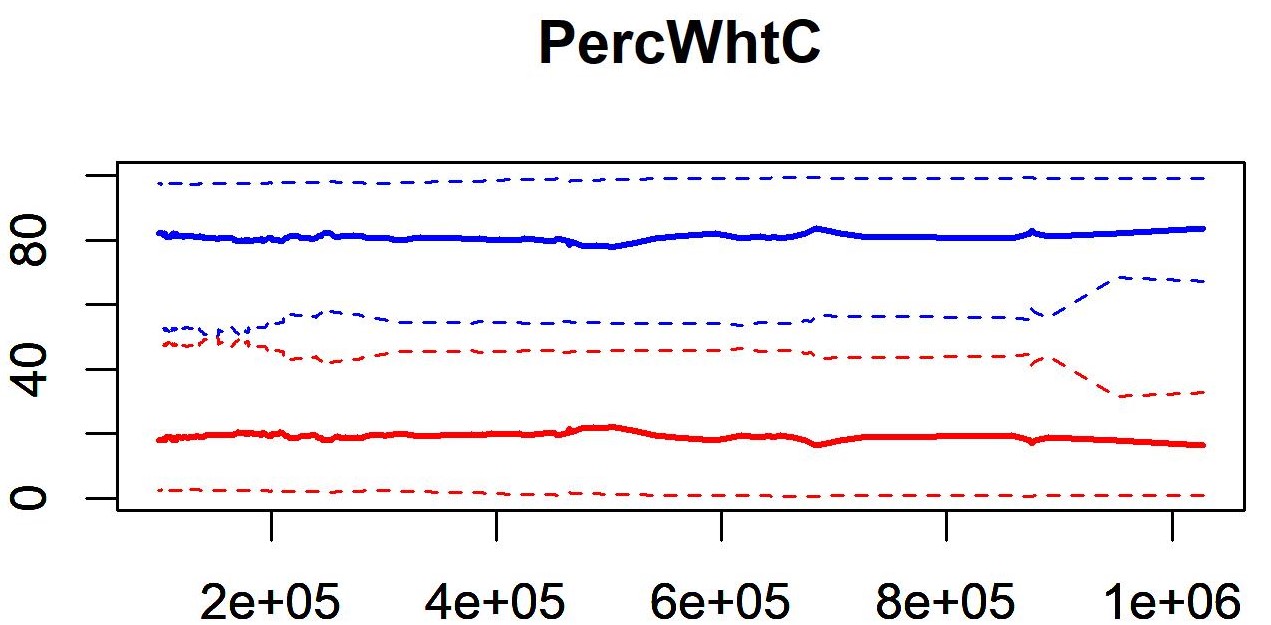}
\caption*{
\footnotesize
Blue = Nondesert areas; red = banking deserts.\\
10, 50, and 90\% quantile values are depicted with dashed/solid/dashed lines for desert and nondesert areas.\\
X-axis = City population.\\
Calculated beginning with N = 10 largest cities and increasing sequentially to include smaller cities until N = 319.  
}
\end{center}
\end{figure}

While large cities (in orange) are shown not to be clustered by desert size or any socioeconomic variable, their bivariate relationships might differ versus small cities or the entire sample. Three separate regressions are therefore calculated for all, large, and small cities. Differences, mainly in the intercept, are visually apparent, while the slope and intercept coefficients are presented in Table 1. The relationship between the percentage White and desert population shares is insignificant for all cities, and there is no relationship between desert size and the percentage of Black residents for the largest cities. Among the three regressions for each pair, most coefficients are very similar, although these small differences are likely to be statistcally significant. Large cities have higher poverty overall, for example, but have a weaker relationship between the poverty rate and desert size, since these cities tend to have larger intercepts and smaller slopes for all variables. 

Next, desert and nondesert areas are compared. The first step is to compare the socioeconomic indicators in each area across city size. Figure 6 shows that most of the quantile values (at 10\%, 50\%, and 90\%) are stable over size. The medians of the percentages are roughly similar in both areas, with one exception: The median percent Black is higher in deserts than nondeserts in cities with between 500,000 and one million residents. The "extreme" 90\% band is wider for this variable, showing that nearly all-Black tracts are much more commonly found in banking deserts.

Nearly all of the concentration quantiles for all three variables higher in are nondeserts; these are more likely to contain the majority of a city's residents in poverty, as well as both Black and White residents. This confirms the idea from Figure 4 that even when defined banking deserts make up part of a city's poorest, least White neighborhoods, similar block groups will be classified as non-deserts. 

\begin{figure}[h]
\begin{center}
\hfill

\caption{Distributions of Socioeconomic Characteristics in Desert and Non-Desert Areas.}
\includegraphics[width=.4\textwidth]{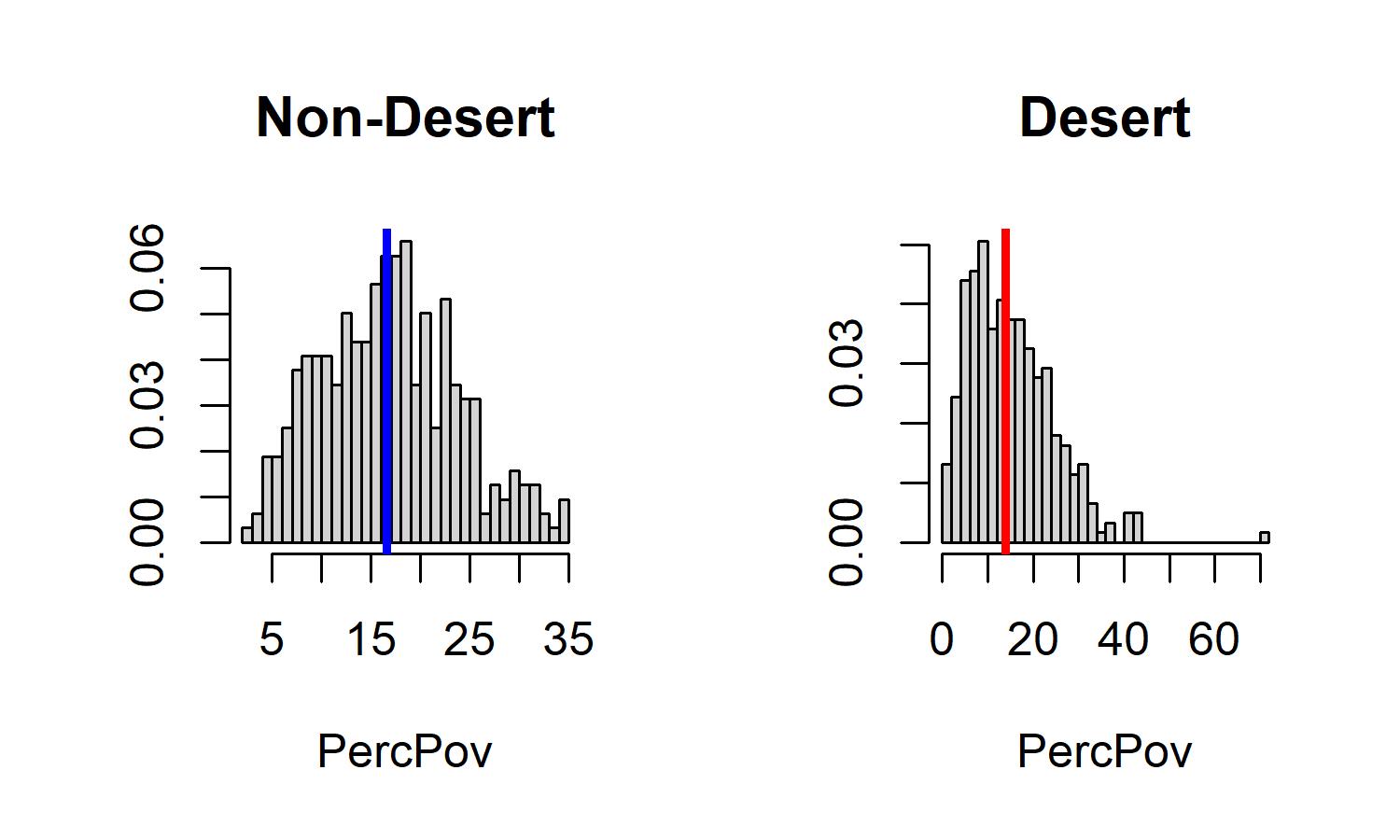}
\includegraphics[width=.4\textwidth]{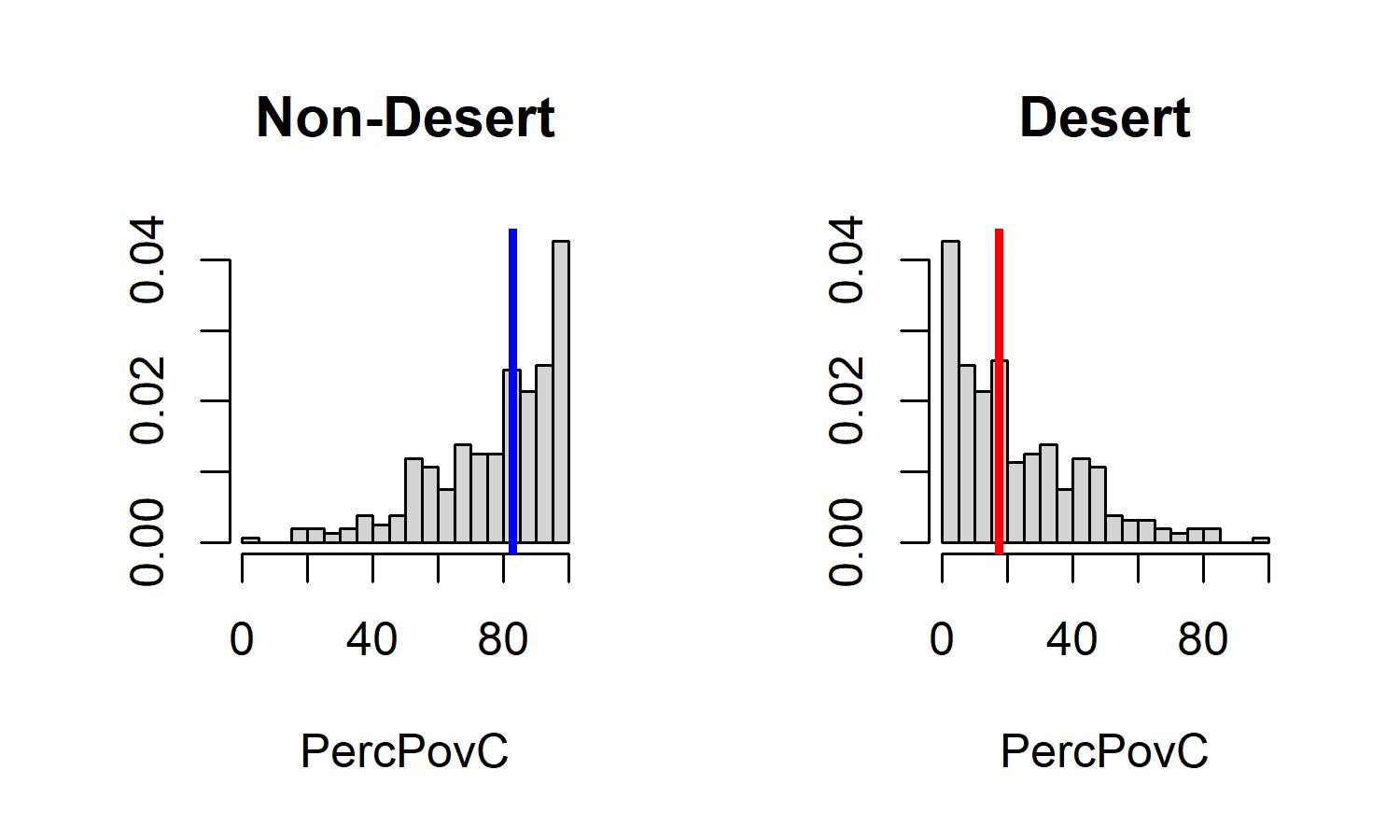}

\includegraphics[width=.4\textwidth]{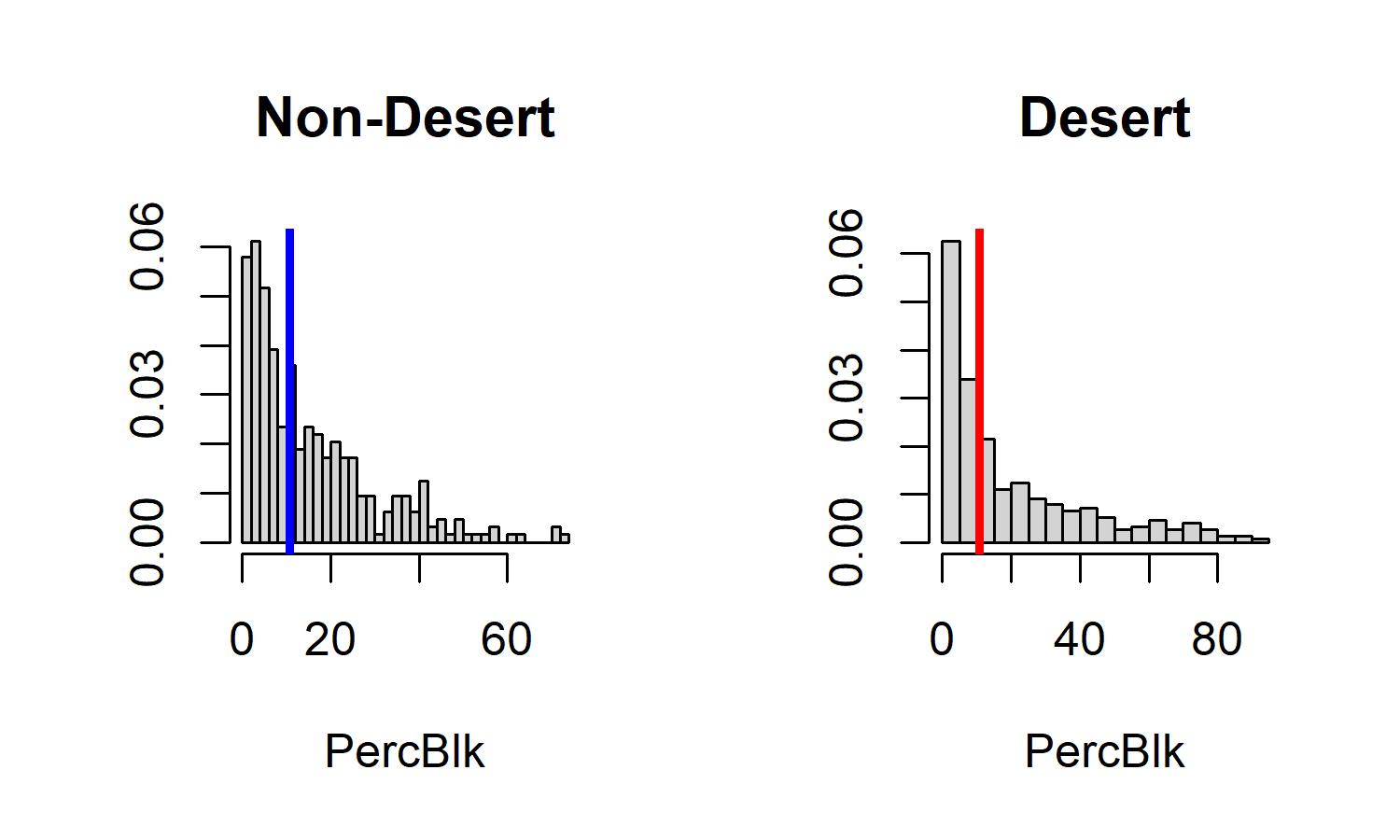}
\includegraphics[width=.4\textwidth]{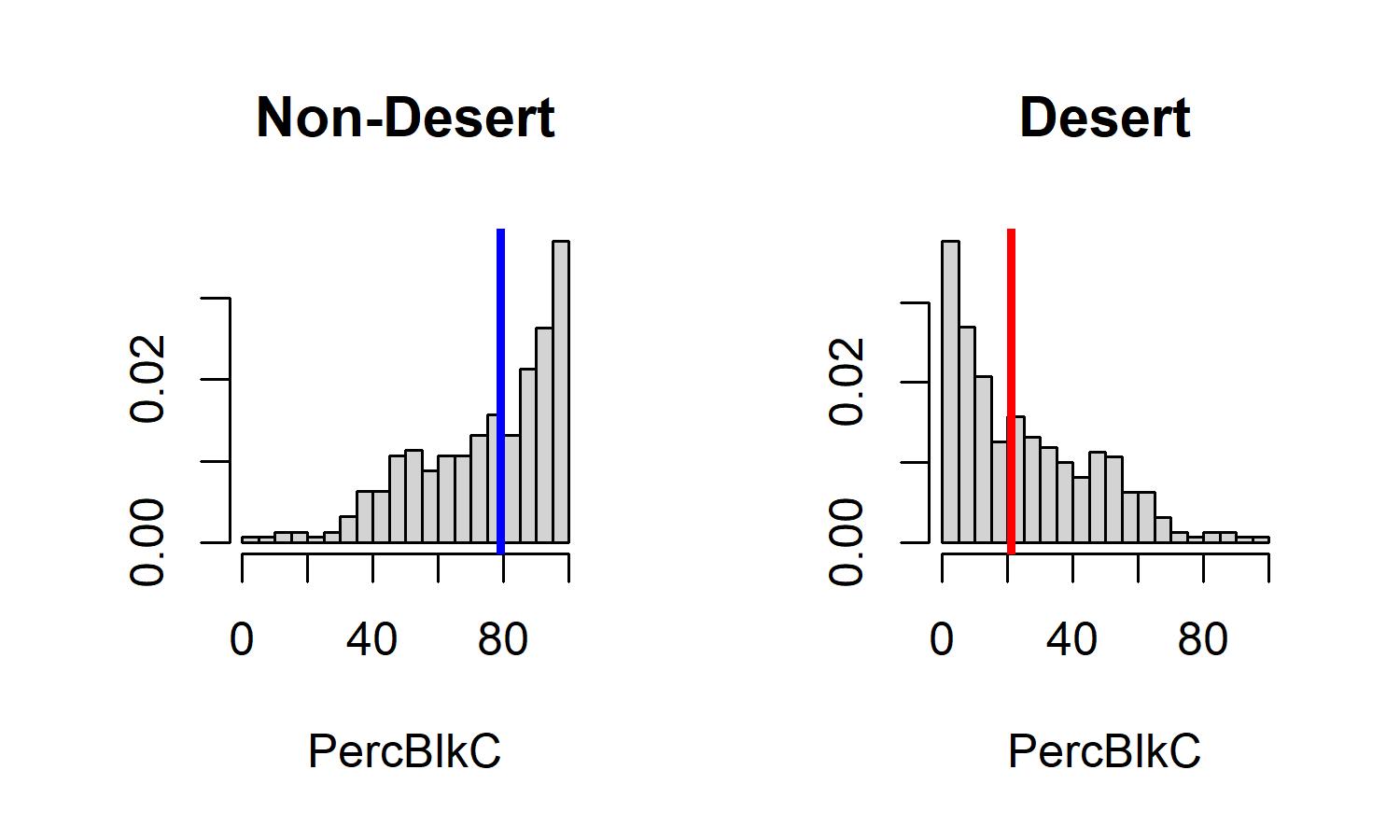}

\includegraphics[width=.4\textwidth]{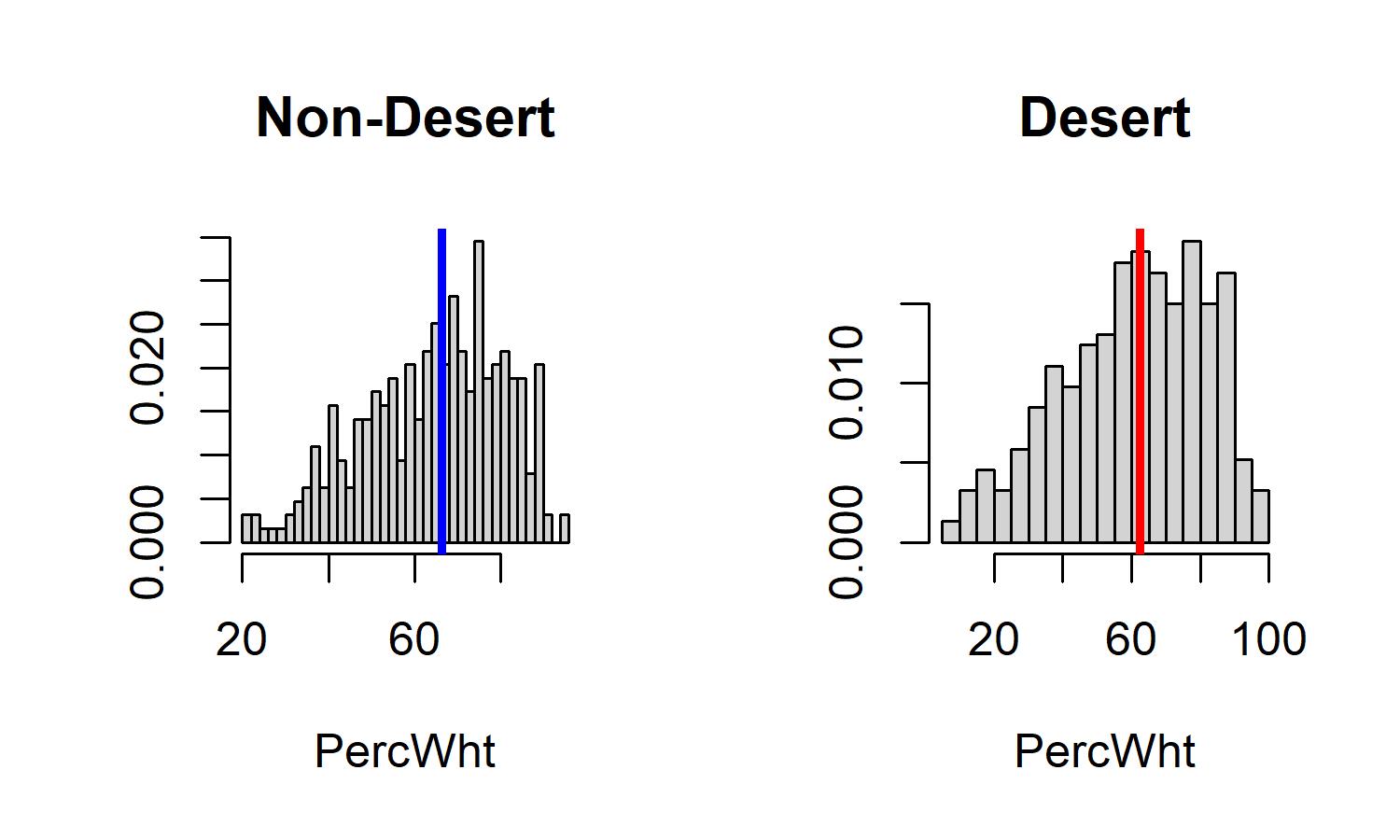}
\includegraphics[width=.4\textwidth]{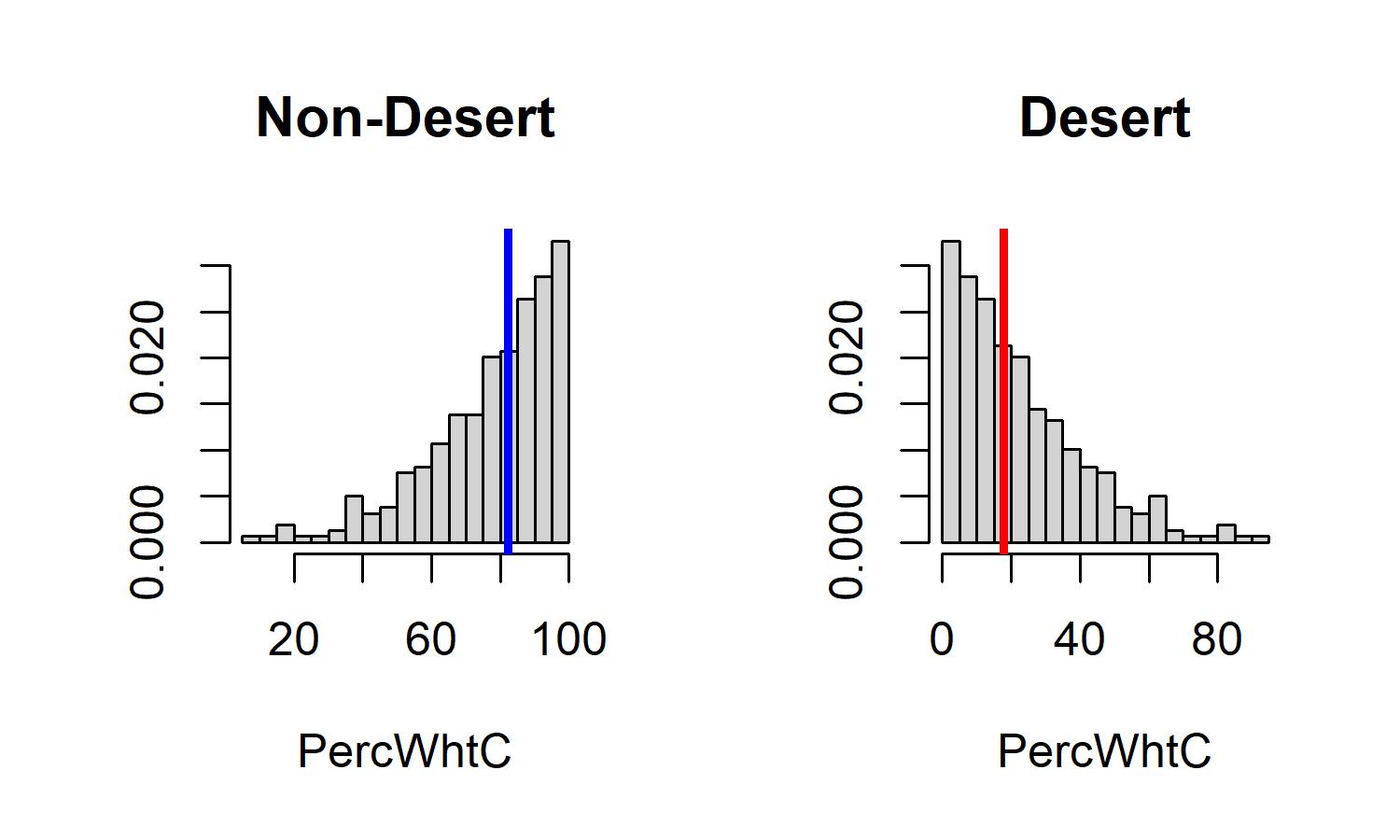}

\caption*{
\footnotesize
Numerical values for each measure calculated separately for desert and non-desert areas within each city.\\
Vertical line = median value for N = 319 cities.\\
See Table 2 for numeric median values and variable explanation: C = "Concentration" of a population in a specified area. For example, \textit{PercPov} = (\# in poverty/area population), while \textit{PerPovC} = (\# in poverty/total number of people in poverty). 
}
\end{center}
\end{figure}

Based on these results, it is possible that the typical city's non-desert areas are poorer and Whiter than the deserts themselves. Figure 7 shows the distributions of the six socioeconomic variables examined in this study. The percentages of Black and White residents have similar distributions in the two areas. While the percentage in poverty has similar medians, the desert distribution skews right, with a disproportionate share of cities with low poverty in these areas. The concentration measure are almost "mirror images" in the two areas; there are many cities with most Whites outside of banking deserts, for example.

\begin{figure}[h]
\begin{center}
\hfill

\caption{Poverty Differentials: Distributions and By City Size.}
\includegraphics[width=.8\textwidth]{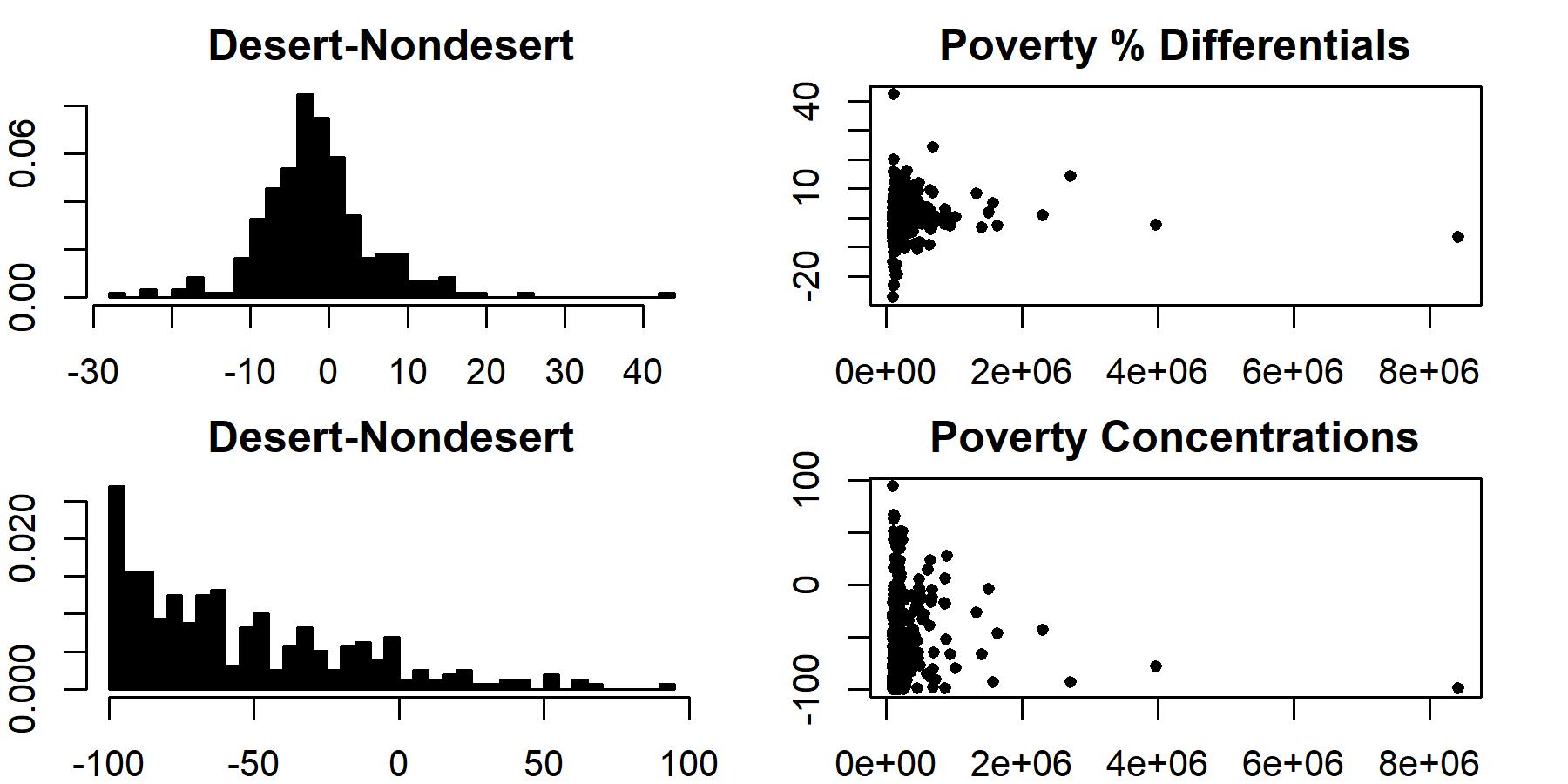}

\caption*{
\footnotesize
Calculated as percentage differences: desert minus nondesert values.\\
Top panel = Differences in poverty rates.\\
Bottom panel = Differences in poverty concentrations.
}
\end{center}
\end{figure}

Figure 8 confirms that deserts tend to have lower poverty rates than other parts of these cities, with the mean and the bulk of the distribution left of the zero line. A plot across city size shows little relationship as well. Likewise, the majority of differences in poverty concentrations are also negative, and again uncorrelated with city size. This again provides support to the notion that banking deserts are not synonymous with poverty. This is likely due to the prevalence of equally poor block groups with more bank access.  

\begin{table}[h]
    \begin{adjustwidth}{-.5in}{-.5in}  
\caption{Quantile Values of Main Variables.}
        \begin{center}
\begin{tabular}{lrrrrrrrr}
Non-desert&10\%&50\%&90\%&Desert&10\%&50\%&90\%\\
\hline
PercPov&7.9&16.6&25.7&PercPov&4.8&13.9&27.68\\
PercPovC&51.46&82.8&98.32&PercPovC&1.68&17.2&48.54\\
PercWht&41.1&66.4&84.2&PercWht&31.54&62.5&87.1\\
PercWhtC&52.76&82.1&97.54&PercWhtC&2.46&17.9&47.24\\
PercBlk&1.9&10.7&37.48&PercBlk&1.1&10.6&54.78\\
PercBlkC&43.96&79&98.44&PercBlkC&1.56&21&56.04\\
\hline
\end{tabular}
\end{center}
\end{adjustwidth}
\caption*{
\footnotesize
Calculated from the N = 319 city-level values for each variable.\\
C = "Concentration" of a population in a specified area. \\For example, \textit{PercPov} = (\# in poverty/area population), while \textit{PerPovC} = (\# in poverty/total number of people in poverty). }
\end{table}

Quantile values for each variable, for deserts and nondesert areas in all 319 cities, are presented in Table 2. One interesting finding is that while the 10\%, median, and 90\% values are relatively similar across both areas, concentrations (for all variables) are much higher in nondeserts. The percentage of Black residents is only higher in banking deserts at the 90\% quantile. This suggests that banking deserts might not be significantly poorer or less White than other parts of these cities.

Table 3 presents the 10\% of cities with the largest and smallest poverty differentials. A future study might model this variable empirically, but one interesting finding is that more large cities have positive differentials (where deserts have higher poverty rates) than nondeserts. Examples include Chicago, Atlanta, and Dallas, while Las Vegas is the largest city in the negative group. The median populations are 233,000 and 130,000, respectively. Since poverty differentials are shown not to be directly related to city size, explanatory variables need to be identified and tested. 

The only regions where the alternative might hold is the U.S. Northeast and possibly the Midwest. Figure 8 shows four separate distributions for (\textit{desert - nondesert}) poverty differentials; cities in the South and West have negative median values. The Midwest median is slightly above zero, with a number of positive values. The median is highest in the Northeast. Further investigation is needed to address these potential regional differences.

\begin{figure}[h]
\begin{center}
\hfill

\caption{Poverty Differentials: Distributions by Region.}
\includegraphics[width=.5\textwidth]{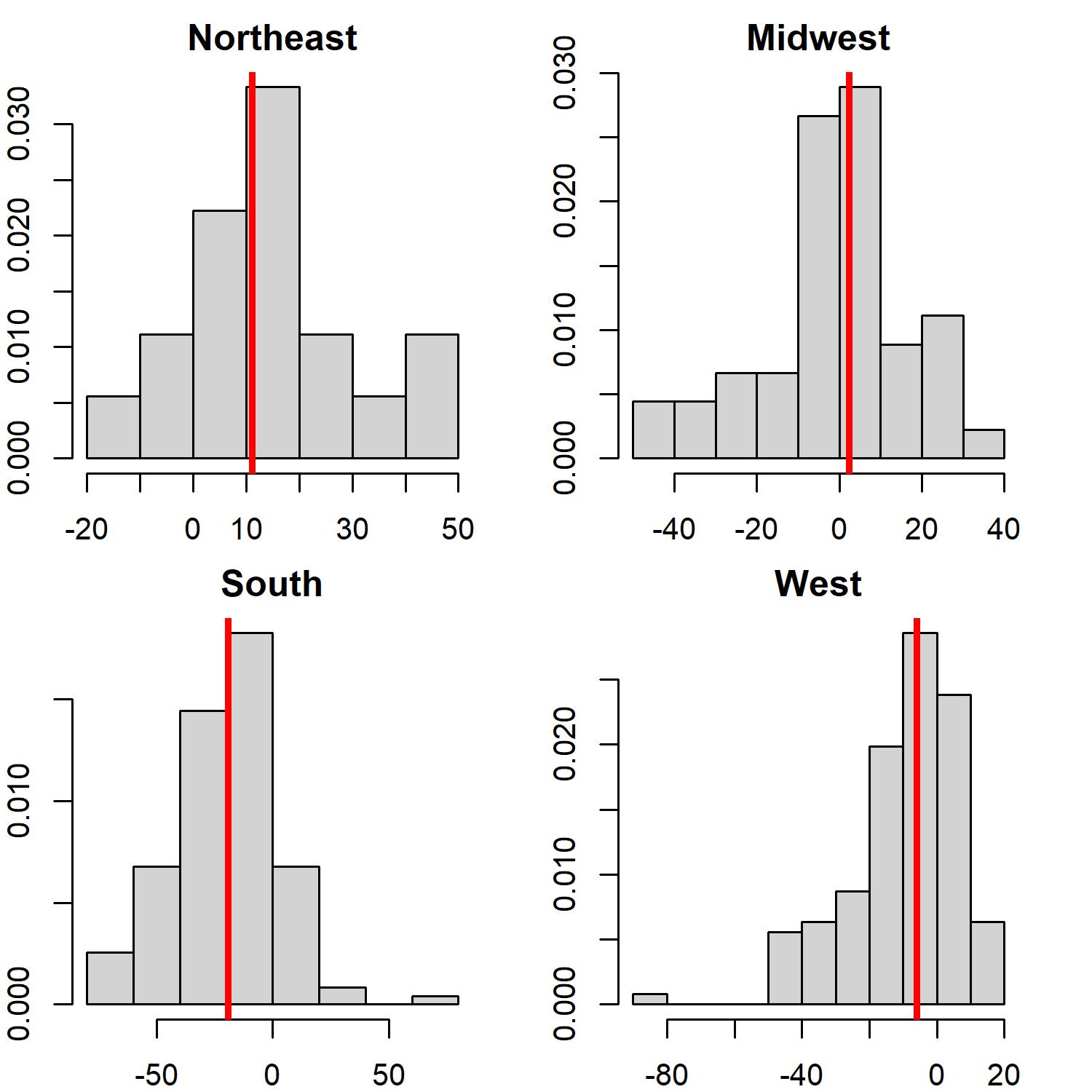}

\caption*{
\footnotesize
Calculated as percentage differences: desert minus nondesert values.\\
N = 24 (Northeast), 47 (Midwest), 119 (South), and 129 (West).\\
Vertical line = median value.
}
\end{center}
\end{figure}

Overall, these results show that there is little statistical difference in of banking deserts in different types of cities. This might be contrary to assumptions that larger cities might have larger deserts. Likewise, when comparing deserts and non-deserts in this sample of cities, there is no evidence that banking deserts are poorer or less White than nondeserts.

\section{Conclusions}
Previous research, analyzing different cities, geographic areas, and spatial scales, has found neighborhoods across the United States that have limited financial access and few bank branches. While there can be severe economic and social consequences resulting from financial exclusion, many of these results have been shown to be mixed in empirical analyses. City-level studies have uncovered wide variation, and national-level studies have found "banking deserts" to be mainly concentrated in central cities. 

This study provides the first comprehensive analysis of medium (more than 100,000 residents) and large (more than 250,000 residents) cities for the entire United States. Contrary to certain prior beliefs, there is little evidence that large cities have poor areas with worse financial access than do small cities; there appears to be no relationship between bank densities and city size or poverty rates.

Classifying block groups as "banking deserts" if they have fewer than one bank in a 1-mile radius of their centroids (or fewer than four within 2 miles), the median desert size in the sample of 319 cities is about 20 percent. The bivariate relationship between desert size and a set of six socioeconomic variables is weaker for the subsample of 83 "large" cities with more than 250,000 residents. The rolling quantile values of these indicators (the rates and concentrations of the poverty rate, the percent Black, and percent White) also change little across size ranges. Only the percent Black is relatively large--the median desert value is equivalent to the highest nondesert values--in cities larger than a half million residents. For the other variables, the nondesert quantiles tend to be higher than desert values; ths might be unexpectied but is probably because many poor, non-White block groups are not classified as "banking deserts."

Only in the U.S. Northeast, and possibly the Midwest, might there be more evidence of a predominance of poor, non-White desert areas.Likewise, there are more large cities where deserts have higher poverty rates in deserts than in nondeserts. But the findings presented here help provide some context in distinguishing between perceptions of "banking deserts" and the fact that poverty is likely a more extensive problem than is geographically determined banking access.

\begin{table}[!bp]
\footnotesize
    \begin{adjustwidth}{-.5in}{-.5in}  
\caption{Largest Poverty Differentials (Deserts vs. Non-Deserts).}
        \begin{center}
\begin{tabular}{lrrrr}
City&Pop&\% Pov (Desert)&\% Pov (Non-Desert)&Difference\\
\hline
College Station, Texas&113686&70.1&27.8&42.3\\
Boston, Massachusetts&684379&43&18.8&24.2\\
Peoria, Illinois&113532&42.5&22.6&19.9\\
Cincinnati, Ohio&301394&41.7&25.6&16.1\\
Rochester, Minnesota&115557&28.1&12.2&15.9\\
Beaumont, Texas&118151&30.8&15.4&15.4\\
Shreveport, Louisiana&192035&36.8&22.2&14.6\\
Baton Rouge, Louisiana&224149&36.5&22.2&14.3\\
Chicago, Illinois&2709534&32.4&18.2&14.2\\
Newark, New Jersey&281054&40.6&27.3&13.3\\
Buffalo, New York&256480&43.3&30&13.3\\
Lafayette, Louisiana&126666&34&21.8&12.2\\
Lubbock, Texas&253851&32&19.9&12.1\\
Fort Wayne, Indiana&265752&26.3&14.3&12\\
Atlanta, Georgia&488800&29&17.1&11.9\\
Tulsa, Oklahoma&402324&28.3&17.2&11.1\\
Sioux Falls, South Dakota&177117&22.3&12.2&10.1\\
Thornton, Colorado&136868&18&8.1&9.9\\
Springfield, Missouri&167051&33.8&24&9.8\\
Memphis, Tennessee&651932&30.6&21&9.6\\
Pompano Beach, Florida&110062&27.6&18.1&9.5\\
Omaha, Nebraska&475862&22.5&13.2&9.3\\
Cleveland, Ohio&385282&40.8&31.5&9.3\\
Amarillo, Texas&198955&22.1&13.3&8.8\\
Washington, District of Columbia&692683&24.2&15.7&8.5\\
North Las Vegas, Nevada&241369&17.8&9.3&8.5\\
Garden Grove, California&173258&22.2&13.8&8.4\\
Dallas, Texas&1330612&24.9&16.5&8.4\\
Oxnard, California&208154&19.6&11.8&7.8\\
Pueblo, Colorado&110841&31.6&23.9&7.7\\
\hline
West Palm Beach, Florida&109767&9.7&18.7&-9\\
Denton, Texas&136195&13.3&22.3&-9\\
Minneapolis, Minnesota&420324&11.8&20.9&-9.1\\
West Valley City, Utah&136009&6.3&15.4&-9.1\\
Escondido, California&151300&8.8&18.1&-9.3\\
Las Vegas, Nevada&634773&7.9&17.3&-9.4\\
Lexington-Fayette, Kentucky&320601&9.5&19&-9.5\\
Glendale, California&200232&5.2&14.8&-9.6\\
Lowell, Massachusetts&111306&9.8&19.5&-9.7\\
Norman, Oklahoma&122837&10.1&20.2&-10.1\\
Round Rock, Texas&124434&0.8&10.9&-10.1\\
Coral Springs, Florida&132568&1.4&11.6&-10.2\\
Santa Maria, California&106224&6.6&16.8&-10.2\\
Lehigh Acres, Florida&123378&19&29.3&-10.3\\
Irvine, California&273157&3&13.7&-10.7\\
Syracuse, New York&142874&21.3&32&-10.7\\
Miami, Florida&454279&12.6&23.4&-10.8\\
Hartford, Connecticut&123088&18.4&29.5&-11.1\\
Clarksville, Tennessee&152934&11.1&22.5&-11.4\\
Gresham, Oregon&110494&7.1&19.2&-12.1\\
El Cajon, California&103186&7.5&22.8&-15.3\\
Palmdale, California&156293&11.9&28.2&-16.3\\
Gainesville, Florida&132127&18.3&34.8&-16.5\\
Fargo, North Dakota&121889&0&16.6&-16.6\\
El Monte, California&115517&4.1&21.1&-17\\
Elizabeth, New Jersey&128333&0&17.7&-17.7\\
Eugene, Oregon&168302&6.3&25.9&-19.6\\
Charleston, South Carolina&135257&4.8&24.6&-19.8\\
Provo, Utah&116403&11.3&34.3&-23\\
Boulder, Colorado&106392&7.3&30.7&-23.4\\
Las Cruces, New Mexico&102102&4.9&32.1&-27.2\\

\hline

\hline
\end{tabular}
\end{center}
\end{adjustwidth}
\end{table}

\end{document}